\RequirePackage{lineno}
\documentclass{nature}
\usepackage{xcolor}

\usepackage{caption}
\usepackage{amsmath}
\usepackage{bm}
\usepackage{braket}
\usepackage{amsmath}
\usepackage{amssymb}
\usepackage{textcomp}
\usepackage{float}

\usepackage{bbold}
\usepackage{mathdots}

\title{Pseudospin-selective Floquet band engineering in black phosphorus}

\author{Shaohua Zhou$^{1,*}$, Changhua Bao$^{1,*}$, Benshu Fan$^1$,  Hui Zhou$^{2}$, Qixuan Gao$^1$, Haoyuan Zhong$^1$, Tianyun Lin$^1$,   Hang Liu$^{2,3}$,  Pu Yu$^{1,4}$, Peizhe Tang$^{5,6}$, Sheng Meng$^{2,3}$, Wenhui Duan$^{1,4}$ and Shuyun Zhou$^{1,4,\dagger}$}

\usepackage{graphicx}
\makeatletter
\let\saved@includegraphics\includegraphics
\AtBeginDocument{\let\includegraphics\saved@includegraphics}
\renewenvironment*{figure}{\@float{figure}}{\end@float}
\makeatother

\begin{document}
\maketitle

\begin{affiliations}

  \item State Key Laboratory of Low-Dimensional Quantum Physics and Department of Physics, Tsinghua University, Beijing 100084, P. R. China
 \item Institute of Physics, Chinese Academy of Sciences, Beijing 100190, P. R. China
 \item Songshan Lake Materials Laboratory, Dongguan, Guangdong 523808, P. R. China
 \item Frontier Science Center for Quantum Information, Beijing 100084, P. R. China
 \item School of Materials Science and Engineering, Beihang University, Beijing 100191, P. R. China
 \item Max Planck Institute for the Structure and Dynamics of Matter, Center for Free Electron Laser Science, 22761 Hamburg, Germany\\
 
 *These authors contributed equally to this work\\
 $\dagger$e-mail: syzhou@mail.tsinghua.edu.cn
\end{affiliations}

\begin{abstract}
{\bf Time-periodic light field has emerged as a control knob for manipulating quantum states in solid-state materials\cite{Oka2009PhotoHE,Demler2011PhotoHE,Lindner2011natphy}, cold atoms\cite{Coldatom2014} and photonic systems\cite{Photonic2013} via hybridization with photon-dressed Floquet states\cite{Shirley1965Floquet} in the strong coupling limit, dubbed as Floquet engineering.   Such interaction leads to tailored properties of quantum materials\cite{okaRev2019,Lindner2020,Chris2021rev,Sentef2021,ZhouNRP2021}, for example, modifications of the topological properties of Dirac materials\cite{WangYH2013,CavalleriNP20} and modulation of the optical response\cite{Wang2014Stark,Gedik2015Stark,Hsieh2021nat}. 
Despite extensive research interests over the past decade\cite{Lindner2011natphy,Podolsky2013PRL,RubioNL2016,DevereauxNC2016,zhang2016theory,Lindner2020}, there is no experimental evidence of momentum-resolved Floquet band engineering of semiconductors, which is a crucial step to extend Floquet engineering to a wide range of solid-state materials.  Here, based on time- and angle-resolved photoemission spectroscopy measurements, we report experimental signatures of Floquet band engineering in a model semiconductor - black phosphorus.  Upon near-resonance pumping at photon energy of 340 to 440 meV, 
a strong band renormalization is observed near the band edges. In particular, light-induced dynamical gap opening is resolved at the resonance points, which emerges simultaneously with the Floquet sidebands.   
Moreover, the band renormalization shows a strong selection rule
favoring pump polarization along the armchair direction, suggesting pseudospin selectivity for the Floquet band engineering as enforced by the lattice symmetry.  Our work demonstrates pseudospin-selective Floquet band engineering in black phosphorus, and provides important guiding principles for Floquet engineering of semiconductors. }
\end{abstract}

\newpage

\renewcommand{\thefigure}{\textbf{Fig. \arabic{figure} $\bm{|}$}}
\setcounter{figure}{0}

Time-periodic light field can induce photon-dressed electronic states  through virtual absorption or emission of photons, which are called Floquet states in analogy to Bloch states in spatially-periodic crystals\cite{AshcroftMermin}. The interaction between Floquet states provides a fascinating pathway to tailor the electronic, symmetric and topological properties of quantum materials dynamically\cite{okaRev2019,Lindner2020,Chris2021rev,Sentef2021,ZhouNRP2021}, for example, controlling the non-equilibrium topological properties of Dirac materials\cite{Rubio2014,Patrick2016PRB,WangzhongPRL2016,Ran2016PRL,CavalleriNP20}, inducing Floquet topological phase in semiconductors\cite{Lindner2011natphy,Podolsky2013PRL,DevereauxNC2016,zhang2016theory}, as well as modulating the optical response\cite{Wang2014Stark,Gedik2015Stark,Hsieh2021nat} and tunneling current\cite{LeeSteady2021}.

The most fundamental physics of the Floquet band engineering lies in inducing the band hybridization between the original Bloch states and the photon-dressed Floquet sidebands. 
Taking a semiconductor as an example (Fig.~1a), a strong coupling between the Bloch bands ($m,n$ = 0 states in the framework of the Floquet theory) and the Floquet sidebands, e.g. $n$ = -1 for the conduction band (CB) and $m$ = 1 for the valence band (VB, see Fig.~1b), could lead to a dynamical gap opening at the crossing points (labeled as $\Delta$ in Fig.~1c).  
Moreover, upon near-resonance pumping, the CB edge with $n$ = -1 is close to the VB edge, which could lead to a stronger interaction.
Such time- and momentum-dependent band renormalization can be further enriched by spin\cite{DevereauxNC2016}, valley\cite{RubioNL2016}, and pseudospin degrees of freedom.

Despite surging research interests over the past decade, so far direct experimental demonstration of momentum-dependent Floquet band engineering has been limited to the topological surface state of Bi$_2$Se$_3$ (ref.\cite{WangYH2013}) due to its unique linear dispersions, which can couple with low-energy photons effectively. For semiconducting WSe$_2$, while the photon-dressed sidebands have been observed\cite{GierzWSe2}, no band renormalization has been detected yet. 
Whether such Floquet band engineering is indeed realistic beyond Dirac materials has remained a long-standing question. Answering this question is important, particularly considering that Floquet band engineering of a semiconductor is a critical step toward inducing transient topological states in topologically trivial materials\cite{Lindner2020}.   Here by using time- and angle-resolved photoemission spectroscopy with mid-infrared (MIR) pumping  (TrARPES, see schematic in Fig.~1d) , we report experimental signatures of Floquet band engineering of a semiconducting black phosphorus  upon near-resonance pumping, which exhibits a strong light polarization dependence indicating novel pseudospin selectivity.

\begin{figure}[H]
	\centering
	\includegraphics[width=16.8 cm]{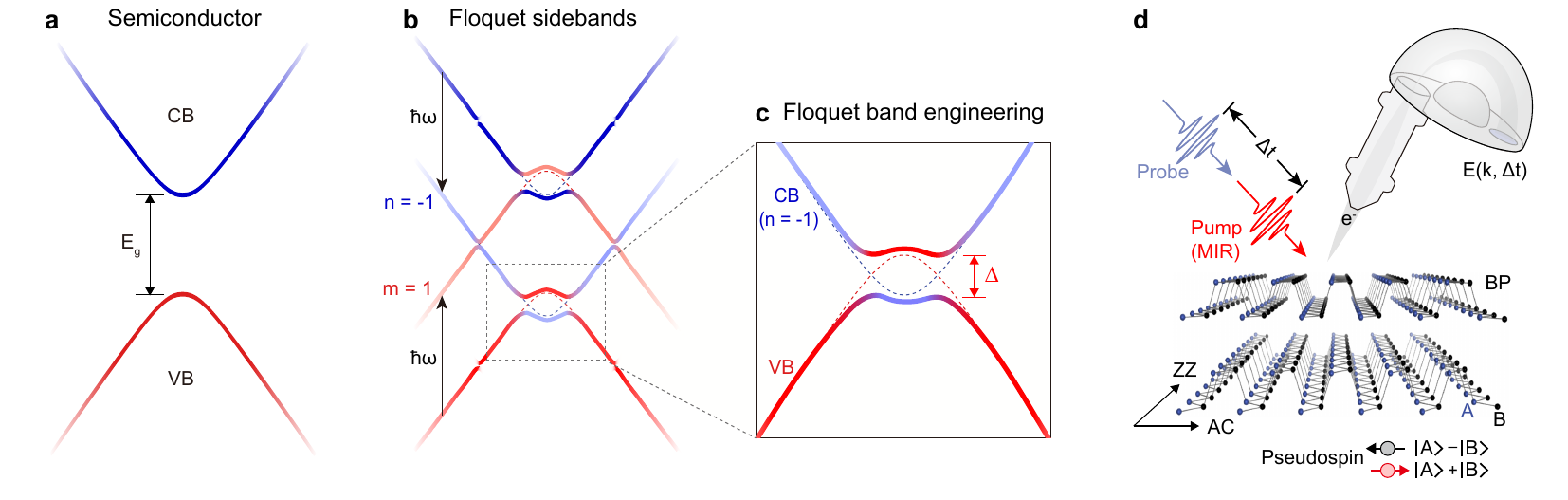}
	\caption{{\bf Schematics for Floquet band engineering in semiconducting black phosphorus.} {\bf a,b}, Schematic dispersions of a semiconductor (\textbf{a}) and corresponding Floquet sidebands (\textbf{b}).  {\bf c}, A schematic of band renormalization between VB and Floquet sideband of CB through Floquet band engineering.  {\bf d}, A schematic for time- and angle-resolved photoemission spectroscopy with mid-infrared pumping on black phosphorus. The inset shows the definition of pseudospin, which is a linear combination of wave functions on the A and B sublattices.}
\end{figure}

\section*{Light-induced band renormalization}

Black phosphorus is a direct band gap semiconductor with anisotropic crystal structure as shown in Fig.~1d.  In contrast to WSe$_2$ with a large band gap of $\sim$ 1.5 eV,  black phosphorus has a smaller band gap of  $E_g\sim$ 0.33 eV, which can be resonantly excited by MIR pump pulses, allowing for efficient Floquet band engineering. 
Moreover, the unit cell contains two sublattices (labeled as A and B in Fig.~1d) which can be viewed as a two-level quantum system in analogy to spin, and their linear superposition defines the pseudospin degree of freedom\cite{Kim2020natmater}.  The coupling of pseudospin with light is strongly anisotropic. In particular, the transition from the VB edge is allowed  only for light polarization along the armchair (AC) direction but not along the zigzag (ZZ) direction, resulting in anisotropic optical properties and strong photoemission intensity modulation\cite{YangLPRB2014,Qiao2014natcom,Cui2015NN,Kim2020natmater}.
Such lattice symmetry enforced pseudospin-selective excitation provides an opportunity to possibly enrich the Floquet band engineering physics with pseudospin selectivity.

\begin{figure}[H]
	\centering
	\includegraphics[width=16.8 cm]{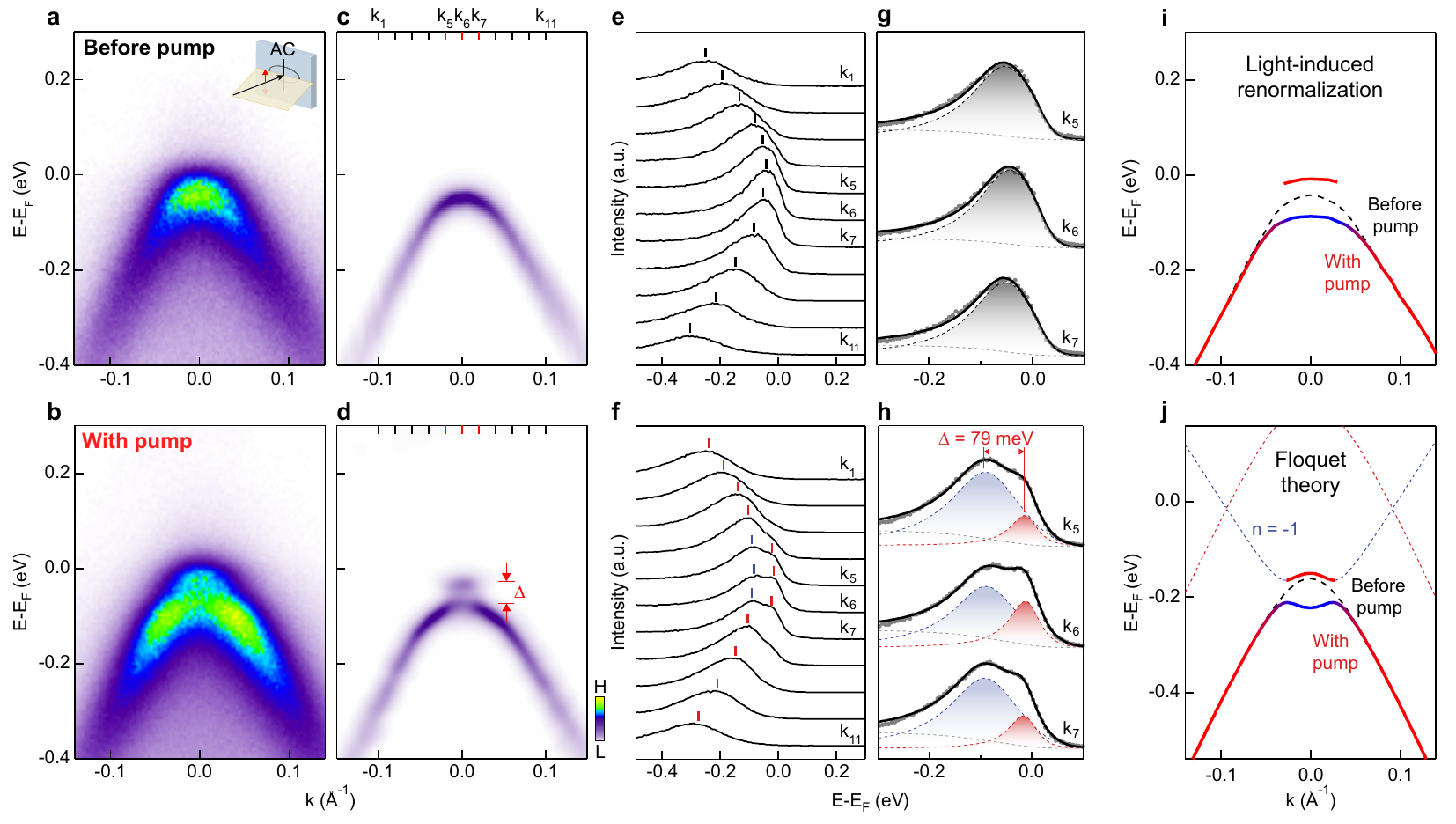}
	\caption{{\bf Observation of light-induced band renormalization.}  {\bf a-d}, TrARPES dispersion images measured along AC direction at $\Delta$t = -1 ps (\textbf{a}) and $\Delta$t = 0 (\textbf{b}), and the corresponding second derivative images (\textbf{c,d}). The pump polarization is along AC direction and perpendicular to the scattering plane ($s$-$pol.$) as shown in the inset in \textbf{a}. The pump photon energy is 380 meV and the pump fluence is 0.7 mJ/cm$\rm^2$. \textbf{e,f}, EDCs for data shown in \textbf{a,b} at momentum points marked by tick marks in \textbf{c,d}.  Red and blue tick marks are the corresponding peak positions. \textbf{g,h}, Fitting of EDCs at resonance points ($k_5$ and $k_7$) and $\Gamma$ ($k_6$) with Lorentzian peaks multiplied by Fermi-Dirac function plus a Shirley background. \textbf{i}, Extracted dispersions before pump (black dashed curve) and with pump (red and blue curves).  \textbf{j}, Calculated VB before pump (black dashed curve) and Floquet band structures (blue and red curves) with pump from the $ab~initio$ tight-binding calculations within the Floquet theory to compare with \textbf{i}. }
\end{figure}

To search for signatures of Floquet band engineering, 
the black phosphorus is pumped above the band gap, so that the Bloch bands and light-induced sidebands overlap  and interact with each other.  Figure 2a,b shows a comparison of dispersion images measured before and upon pumping at 380 meV photon energy, which is slightly above the band gap.  The pump polarization is along the AC direction of the sample and perpendicular to the scattering plane ($s$-$pol.$, see Extended Data Fig.~1 for the TrARPES experimental geometries). Since there is no electric field component perpendicular to the sample surface to couple with the photoemission final states, laser-assisted photoemission or Volkov states\cite{mahmood2016} (photon-dressed photoemission final states) can be excluded. A striking modification of the electronic structure is observed near the VB edge upon pumping. In particular, the single parabolic-like VB in the equilibrium (Fig.~2a) changes into two bands with a hat-like shaped dispersion (Fig.~2b).  Such change is more pronounced by taking the second derivative of the dispersion images (Fig.~2c,d, see Methods for details), which is a common practice to visualize the ARPES dispersion directly.  The dispersions can be extracted by fitting the energy distribution curves (EDCs) in Fig.~2e-h. A comparison of the extracted dispersions before pumping (black dashed curve  in Fig.~2i) and upon pumping (red and blue curves) shows that there is clearly a light-induced band renormalization in black phosphorus.

The light-induced band renormalization observed under the $s$-$pol.$ pumping geometry can be explained by Floquet band engineering, namely, the interaction between VB and Floquet sideband of CB with $n$ = -1 leads to a hybridization gap near the resonance points. The hybridization gap is extracted to be 79 $\pm$ 20 meV from the peak separation at the resonance points $k_5$ and $k_7$ in Fig.~2h. 
The band renormalization is further supported by $ab~initio$ tight-binding calculation within the Floquet theory, which shows a renormalized electronic structure for the Floquet states upon pumping  (Fig.~2j, see Methods for more details).  
We note that the observed light-induced band renormalization is in contrast to previous TrARPES studies\cite{Nurmamat2018bp, chen2018BP, roth2019p2Dmater,Perfetti2D,Carpene2021,Monney2021} with pump  photon energy around 1.5 eV, highlighting the importance of pumping  at previously unexplored MIR regime\cite{Oka2009PhotoHE,Sentef2021,ZhouNRP2021} for Floquet engineering.

\section*{Evidences for Floquet band engineering}

\begin{figure}[H]
	\centering
	\includegraphics[width=16.8 cm]{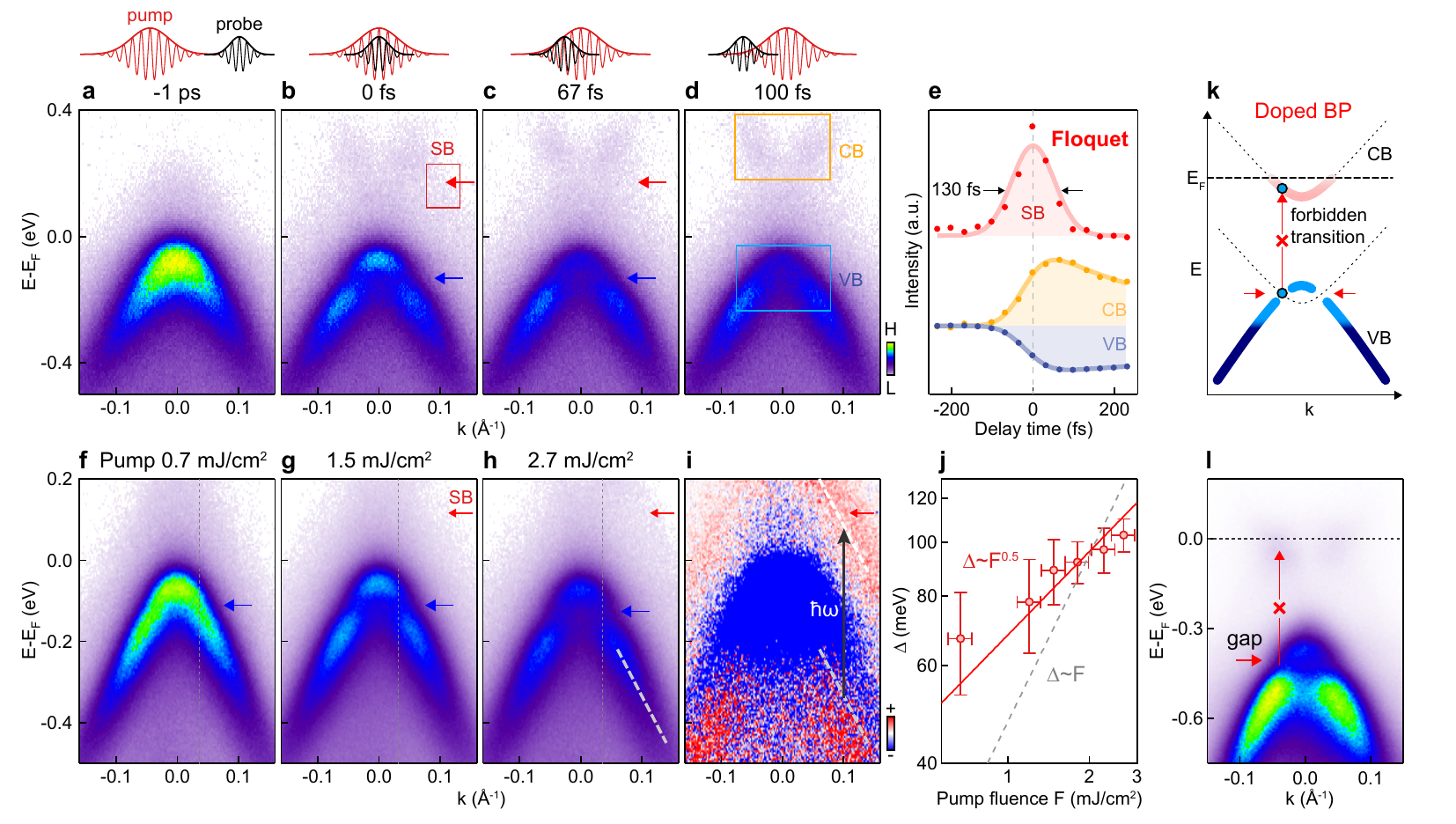}
	\caption{{\bf Supporting evidence of Floquet band engineering.}  {\bf a-d}, TrARPES dispersion images measured  at different delay times along AC direction. The pump photon  energy is 420 meV and the pump fluence is 2.3 mJ/cm$\rm^2$.  The pump polarization is parallel to AC direction. The blue and red arrows point to the hybridization gap and Floquet sideband respectively. {\bf e}, Intensity of the Floquet sideband (integrated over red box in  \textbf{b}), CB and VB (integrated over orange and blue boxes in \textbf{d}) as a function of delay time. {\bf f-h}, TrARPES dispersion images measured  at $\Delta$t = 0 with increasing pump fluence. {\bf i}, Differential image obtained by subtracting data taken at $\Delta$t = -1 ps from \textbf{h}.  The gray and white lines are guides to the VB and Floquet sideband. {\bf j}, Extracted hybridization gap as a function of the pump fluence. {\bf k}, Schematic dispersion for Floquet gap opening of electron-doped black phosphorus upon near-resonance pumping. {\bf l}, TrARPES dispersion images measured at $\Delta$t = 0 in electron-doped black phosphorus with pump photon energy of 400 meV and pump fluence of 3 mJ/cm$\rm^2$.}
\end{figure}

The Floquet band engineering is supported by the same temporal evolution of the light-induced renormalization and Floquet states. 
Figure 3a-d shows a series of snapshots of the transient electronic structures measured at different delay times. The sideband, which is shifted by the pump photon energy, is observed  at a higher pump fluence in the $s$-$pol.$ pump scattering (pointed by red arrows in Fig.~3b,c), suggesting that it is contributed by pure Floquet states\cite{mahmood2016}.  The temporal evolution of the Floquet sideband intensity (red symbols in Fig.~3e) shows that it exists only near time zero with a resolution-limited lifetime of 130 fs.
Moreover, the light-induced band renormalization is also observed in the same timescale  near time zero (pointed by blue arrows in Fig.~3b,c) and becomes barely detectable at later delay time, e.g. 100 fs (Fig.~3d).  Such similar timescale for the Floquet states and light-induced band renormalization implies that the dynamical change in the electronic structure is probably induced by the Floquet band engineering.

The Floquet band engineering is further supported by the pump fluence dependence of the renormalization gap. 
Figure 3f-h shows a few representative TrARPES dispersion images measured at $\Delta$t = 0 with different pump fluences, and the Floquet sideband is observed in the dispersion images (red arrow in Fig.~3g,h) and the differential image (Fig.~3i).  With increasing pump fluence, the dispersion shows a stronger band renormalization with systematically enhanced hybridization gap (Fig.~3j, see Extended Data Fig.~2 for more data and analysis).  The extracted hybridization gap $\Delta$ scales with the pump fluence $F_{pump}$ by $\Delta \propto F_{pump}^{0.5}$. This is in agreement with the expected scaling in the framework of Floquet engineering\cite{Lindner2020} as demonstrated in Bi$_2$Se$_3$ (ref.\cite{mahmood2016}), thereby providing additional support for the Floquet band engineering  of black phosphorus. 

The light-induced renormalization is observed at resonance points where the VB overlaps with the CB with $n$ = -1, which also correspond to where direct optical transition from VB to CB occurs. We note that photoexcited holes left in the VB, i.e., depleted charges before intraband scattering or electron thermalization sets in, could lead to suppression of intensity at the resonance points, which apparently behaves similar to gap opening.  In order to test this alternative explanation, we have intentionally doped the black phosphorus sample to fill the CB, so that direct optical transition is blocked upon near-resonance pumping (see schematic illustration in Fig.~3k).  In this case, light-induced renormalization is still observed (Fig.~3l, see more detailed analysis in Extended Data Fig.~3).  The observation of similar light-induced renormalization even when direct optical transition is forbidden indicates that it is not caused by charge depletion, but rather by Floquet band engineering.

\section*{Pseudospin-selective Floquet engineering}

\begin{figure}[H]
	\centering
	\includegraphics[width=16.8cm]{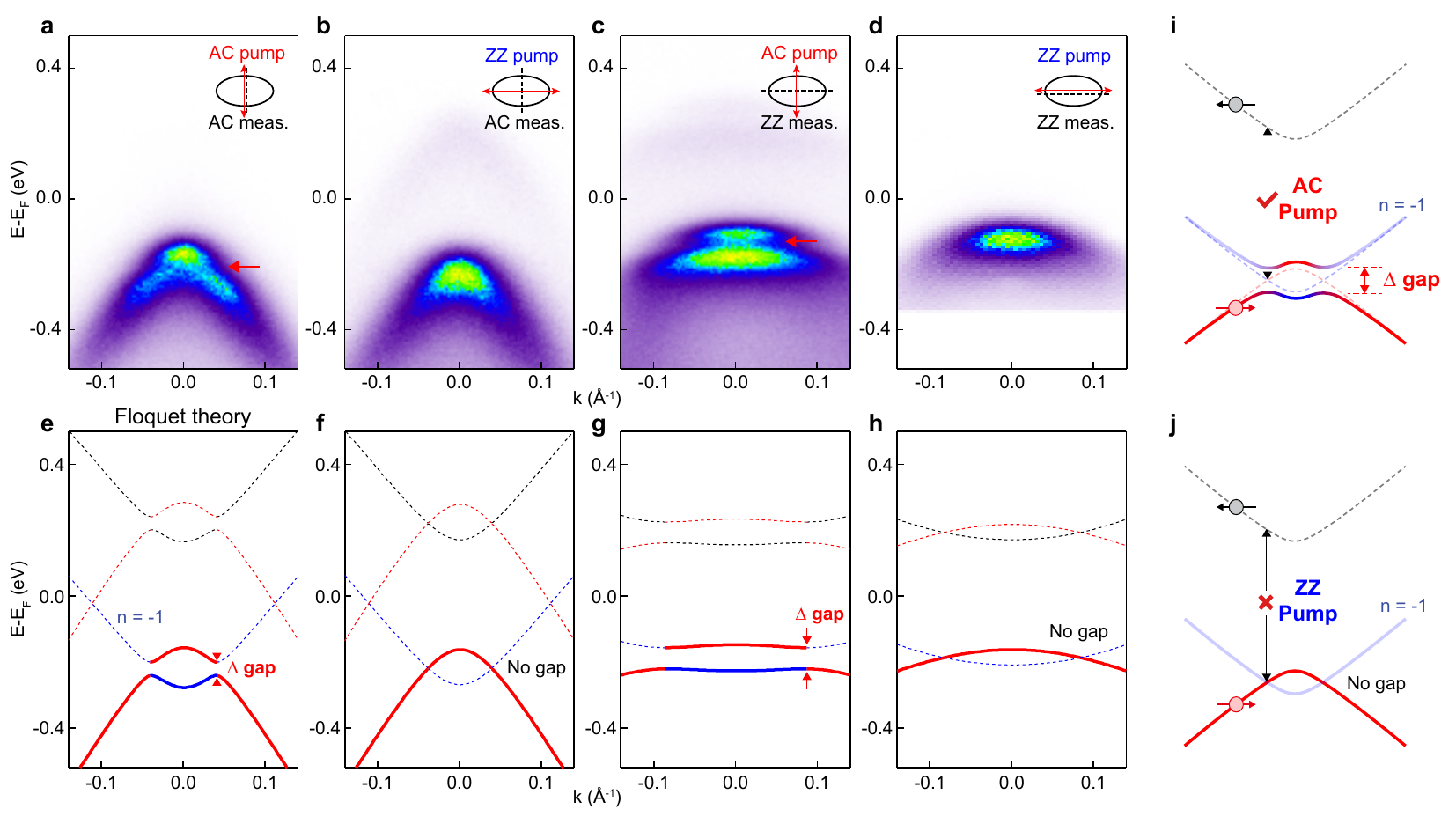}
	\caption{{\bf Pseudospin-selective Floquet band engineering.}  {\bf a-d}, TrARPES dispersion images measured at $\Delta$t = 0 with combinations of different measurement directions and pump polarizations as schematically shown in the insets.  The pump photon energy is 440 meV for {\bf a,b} and 380 meV for {\bf c,d}, and the pump fluence is 0.7 mJ/cm$\rm^2$.  \textbf{e-h}, Calculated Floquet band structures corresponding to \textbf{a-d}. \textbf{i,j}, Schematic summary of pseudospin-selective Floquet band engineering. }
\end{figure}

The light-induced Floquet band engineering presented above is based on TrARPES data measured along AC direction, with pump polarization also along AC direction. To check whether there is any lattice symmetry enforced pseudospin selectivity, we show TrARPES dispersions measured  at four different experimental geometries in Fig.~4 (see corresponding experimental geometry in Extended Data Fig.~1).  Figure 4a,c shows a comparison of TrARPES dispersion images measured with AC pump polarization, and the scattering geometries are $s$-$pol.$ and $p$-$pol.$ respectively. While a stronger sideband is observed for $p$-$pol.$ pump in Fig.~4c due to the interference between Floquet states and Volkov states (``Floquet-Volkov states''\cite{mahmood2016}), the light-induced band renormalization is observed not only for $p$-$pol.$ pump scattering geometry, but also for $s$-$pol.$ pump polarization in which there is no contribution from Volkov states, suggesting that it is independent of the Volkov states (see a summary of the comparison in Extended Data Fig.~4). 
This suggests that the light-induced band renormalization is a fundamental feature to distinguish the Floquet band engineering from the emergence of pure Volkov states, because Volkov states do not renormalize the electronic structure of the host quantum materials while the Floquet engineering does. 
Furthermore, a comparison of Fig.~4a,c and Fig.~4b,d shows that when the pump polarization changes from AC pump to ZZ pump, the renormalization is strongly reduced, and the VB dispersion is similar to the equilibrium state (Fig.~4b,d). These results suggest that the light-induced band renormalization is intrinsically related to the pump polarization with respect to the crystal orientation, namely, there are lattice symmetry enforced pseudospin selection rules.

In the Floquet theory, the light-matter interaction renormalizes the electronic structures of black phosphorus, and therefore the effect of Floquet band engineering is determined by the interaction matrix elements  between electronic states with different Floquet indexed ($m$, $n$) bands.  Theoretical analysis and calculation show that the VB edge with $m$ = 0 and the CB edge with $n$ = -1 are always coupled to open a hybridization gap for AC pump, which becomes significantly reduced for ZZ pump (see calculated results in Fig.~4e-h, Extended Data Fig.~5 and Supplementary Information).  This is in good agreement with TrARPES experimental results and reflects the coupling of light with the pseudospin degree of freedom.
Such pseudospin-selective Floquet band engineering is schematically summarized in Fig.~4i,j, and analogous to the valley-selective Floquet band engineering proposed in monolayer WS$_2$ and WSe$_2$\cite{Gedik2015Stark,DevereauxNC2016,RubioNL2016}. Here we provide experimental signatures of Floquet band engineering in black phosphorus, and show that the pseudospin can indeed play an important role through the light-matter interaction matrix elements (see Supplementary Information for more details).

\section*{Near-resonance versus off-resonance pumping}

\begin{figure}[H]
	\centering
	\includegraphics[width=16.8cm]{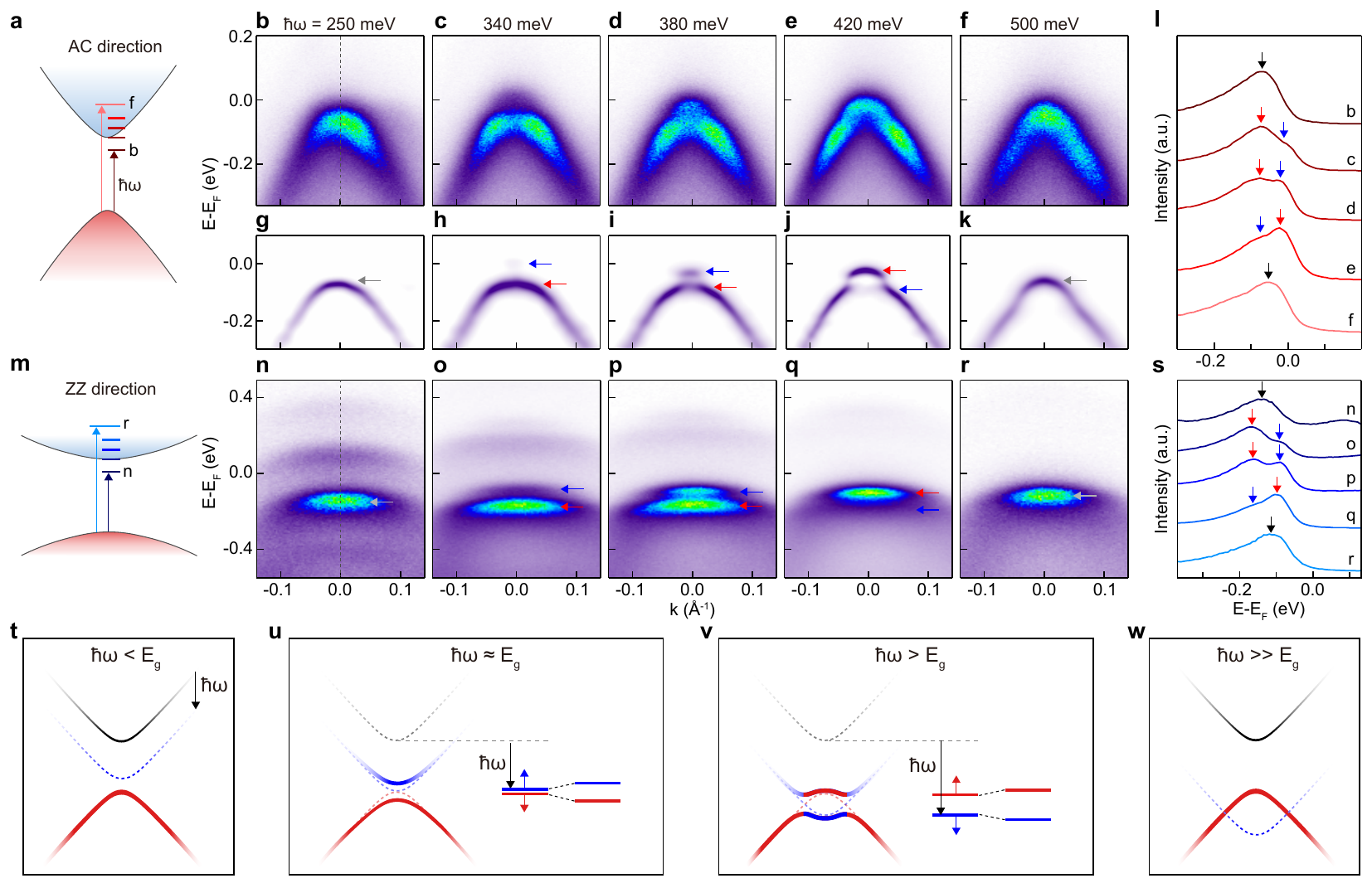}
	\caption{{\bf Evolution of Floquet band engineering with pump photon energy.} \textbf{a}, A schematic for the band structure along AC direction with various pump photon energies used. {\bf b-k}, TrARPES dispersion images measured at $\Delta$t = 0 along AC direction with different pump photon energies ({\bf b-f}) and the corresponding second derivative images ({\bf g-k}). The pump polarization is along AC direction of the sample and $s$-$pol.$ with respect to the scattering plane, and the pump fluence is 0.7 mJ/cm$\rm^2$.    \textbf{l}, EDCs at $k$ = 0 for data shown in \textbf{b-f}. 
	\textbf{m}, A schematic for the band structure along ZZ direction and pump photon energies used.
	{\bf n-r}, TrARPES dispersion images measured at $\Delta$t = 0 along ZZ direction with different pump photon energies.  The pump polarization is along AC direction of the sample and $p$-$pol.$ with respect to the scattering plane. The pump fluence is 0.7 mJ/cm$\rm^2$.  \textbf{s}, EDCs at $k$ = 0 for data shown in \textbf{n-r}.   {\bf t-w}, A schematic summary of electronic structure upon pumping at different photon energies. The insets in \textbf{u},\textbf{v} show the energy shifts for VB edge and CB edge with $n$ = -1.}
\end{figure}

The effects of  light-induced band renormalization upon near-resonance and off-resonance pumping are revealed by changing the pump photon energy across the band gap as schematically illustrated in Fig.~5a. Figure 5b-f shows dispersion images measured along AC direction at different pump photon energies, and the corresponding second derivative images are shown in Fig.~5g-k. 
A clear light-induced band renormalization is observed upon pumping at photon energies from 340 meV to 420 meV (Fig.~5c-e), which are near resonance with the band gap of 330 meV.  Additionally, the transfer of spectral intensity from a stronger lower band to a stronger upper band (pointed by arrows in Fig.~5h-j, and EDCs at the $\Gamma$ point in Fig.~5l) suggests a possible band inversion between the VB edge and the CB edge with $n$ = -1 upon increasing the pump photon energy.  Interestingly, for off-resonance pumping, e.g. 250 meV or 500 meV pump photon energy, the band renormalization is strongly reduced and becomes barely detectable experimentally (Fig.~5b,f,g,k).  Similar pump photon energy dependent light-induced band renormalization has also been observed for dispersions measured along the ZZ direction with AC pump polarization (see Fig.~5m-s). Although there is a stronger sideband due to the Floquet-Volkov states in the $p$-$pol.$ pump scattering geometry, the light-induced band renormalization is also observed upon near-resonance pumping (Fig.~5o-q) and becomes strongly reduced for off-resonance pumping (Fig.~5n,r).  
Such pump photon energy dependent light-induced band renormalization is also supported by theoretically calculated Floquet band structures based on the $ab~initio$ tight-binding Hamiltonian (Extended Data Fig.~6, 7 for a comparison between experimental results and calculated electronic structures). 

Figure 5t-w shows a schematic summary of the pump photon energy dependent Floquet band engineering. For below-gap excitation when the Floquet sideband is far away from the VB, no significant band renormalization is observed. While high-order interaction ($|m-n|>1$) can in principle also lead to light-induced band renormalization, however, the interaction strength is expected to be significantly  reduced\cite{Oka2009PhotoHE}.  Increasing the pump photon energy to near resonance with the band gap, the CB edge with $n$ = -1 approaches the VB edge, leading to significant renormalization as schematically illustrated in Fig.~5u. Further increasing the pump photon energy so that the Floquet sideband overlaps with the VB, a hybridization gap occurs near the resonance points as schematically shown in Fig.~5v.  At even higher pump photon energy, the band renormalization is reduced (Fig.~5w).

The  light-induced band renormalization in black phosphorus is in line with the optical Stark effect\cite{TownesStark}, which is also strongly related to Floquet engineering.  While the optical Stark effect often involves atomic energy levels in atomic physics, Floquet band engineering refers to the dynamical momentum-dependent electronic structure hybridization between the Bloch band and Floquet sideband in the context of condensed matter physics. Here we resolve not only the energy shifts, but also the momentum-dependent band renormalization. Such direct observation of Floquet band engineering in a layered quantum material provides the electronic structure counterpart of the optical Stark effect.

\section*{Discussions and perspectives}

The light-induced band renormalization of black phosphorus reported here distinguishes from previous Dirac materials\cite{WangYH2013,CavalleriNP20} in two major aspects. First, while Floquet band engineering of Bi$_2$Se$_3$ and graphene does not depend on the crystal orientational due to the overall isotropic electronic structure\cite{WangYH2013,mahmood2016}, the Floquet band engineering of black phosphorus exhibits a stronger pump polarization selectivity with respect to the crystal orientation, which is enforced by the lattice symmetry, thereby providing a control knob for turning on and off the Floquet band engineering through the pseudospin degree of freedom. Secondly, while Dirac fermions can couple to low-energy photons at different energies effectively due to the conical dispersion, in semiconducting black phosphorus where the dispersion shows a parabolic behavior near the gap edge, the interaction between the Bloch band and Floquet sideband is strongly enhanced upon resonant pumping near the band gap.

Interestingly, resonance pumping of semiconductors has been conventionally regarded as activating more excitation and scattering channels such as electron-electron and electron-phonon interactions, which might heat up the samples\cite{DevereauxNC2016} and even destroy the Floquet states\cite{GierzWSe2}. 
Our results show that the Floquet states survive within the duration of the pump pulse, and the band renormalization is enhanced upon near-resonance pumping, thereby highlighting the importance of resonance pumping in the Floquet band engineering of black phosphorus. 
Such pseudospin-selective Floquet band engineering offers an exciting opportunity to manipulate time-resolved optical response in black phosphorus via linear dichroic pumping near resonance. 
Finally, we expect that such near-resonance pumping strategy can be applied to more quantum materials, paving a critical step toward the dynamical engineering of the transient electronic structures on the ultrafast timescale, as well as the experimental realization of exotic electronic states such as the Floquet topological phases\cite{Lindner2011natphy,Lindner2020,Sentef2021}.

\renewcommand{\thefigure}{\textbf{Extended Data Figure \arabic{figure} $\bm{|}$}}
\setcounter{figure}{0}

\begin{methods}

\subsection{Sample preparation and electron doping}
~\\
	Black phosphorus single crystals were grown by chemical vapour transport method. A mixture of red phosphorus lump (Alfa Aesar, 99.999\%), tin grains (Aladdin, $\geqslant$99.5\%), and iodine crystals (Alfa Aesar, 99.9\%) was sealed under vacuum in a silica tube. The tube was heated to 400 \textcelsius{} within 2 hours and maintained at 400 \textcelsius{} for 2 hours, then heated to 600 \textcelsius{} and maintained at  600 \textcelsius{} for 1 day. The tube was slowly cooled to 350 \textcelsius{} from 600 \textcelsius{} at a cooling rate of 10 \textcelsius/hour, and then furnace-cooled to room temperature. Millimeter size, high-quality black phosphorus single crystals were obtained. 

	The samples were cleaved and measured at a temperature of 80 K in ultra-high vacuum chamber with a base pressure better than 5 $\times \text{10}^{-\text{11}}$ Torr. The electron-doped black phosphorus is obtained by {\it in situ} deposition of Cs using a commercial dispenser (SAES).  The dispenser was operated at the current of 4.7 A, and the distance between the sample and the source  is $\sim$ 13 cm.  The dispersion is monitored by {\it in situ} ARPES measurements until sufficient electron doping is reached.

\subsection{TrARPES measurements}
~\\
	TrARPES measurements were performed in the home laboratory at Tsinghua University with a regenerative amplifier laser with center wavelength at 800 nm and 10 kHz repetition rate. The pulse energy is 1.3 mJ and the majority of the beam (80$\%$) is used to drive the optical parametric amplifier (OPA). The MIR pulses are produced by non-collinear differential frequency generation (NDFG) using the signal and idle of the OPA. The probe beam with photon energy of 6.2 eV is generated by a three-step fourth harmonics generation process using BBO crystals.  For undoped sample, the Fermi energy is calibrated by the Fermi-Dirac fitting to the spectrum of polycrystalline gold which is electrically connected to the sample holder. For electron-doped sample, there is an overall shift of the electronic structure induced by surface photovoltaic effect with a long relaxation timescale\cite{Perfetti2D,Carpene2021,Monney2021}, and the Fermi energy is obtained by the Fermi-Dirac fitting to the spectrum before pump. 
	
	The pump and probe beams are almost collinear, with a similar incident angle of $\theta$ = 54$^{\circ}$ as shown in Extended Data Fig.~1a.  Four different experimental geometries with different pump polarizations and measurement directions are shown in Extended Data Fig.~1.  The typical pump fluence of 0.7 mJ/cm$^2$ used $s$-$pol.$ pump corresponds to an electric field strength of 6.8$\times 10^7$ V/m. This is calculated from the pump fluence $F$ by $E=t_s\sqrt{2\frac F {\tau}\sqrt{\frac {\mu_0} {\epsilon_0}}}$, where $t_s$ is Fresnel transmission coefficients ($t_s$ = 0.29 for the incident angle of 54$^{\circ}$, calculated from the reflectance in ref.\cite{Cui2015NN} using the method from ref.\cite{WangYH2013}), and $\tau$ is the estimated pump pulse duration (90 fs).

Regarding the penetration depth of the pump and probe beams in TrARPES measurements, the penetration depth of MIR pump light (e.g. 400 meV) is estimated to be 14 nm from the optical absorbance of 4$\%$ per layer\cite{YangLPRB2014}, which is much larger than the probing depth of 3.5 $\pm$ 2 nm by 6.2 eV probe laser estimated from the universal curve\cite{SeahSurf1979_}.  Therefore, the probed beam probes an overall uniformly illuminated region. Regarding the electron-doped sample by deposition of Cs, the depth of electron doping in black phosphorus is estimated to be 15 nm at 80 K\cite{Carpene2021}, which is much larger than the ARPES probing depth. Therefore, TrARPES signals on electron-doped sample are dominated by the electron-doped region near the surface.

A convenient method to directly visualize the dispersion is to extract the peak positions by taking second derivative of the dispersion image, e.g. Fig.~2c,d. Here, a modified second derivative (or 2D curvature) is applied, 
$$\nabla^2I_{ARPES}=\frac{\partial^2 I_{ARPES}}{\partial x^2}+\frac{\partial^2 I_{ARPES}}{\partial y^2}$$
Considering the different units of $k$ and $E$ by taking a transformation for x and y, this becomes
$$C(x,y)\sim\frac{[1+C_x(\frac{\partial f}{\partial x})^2]C_y\frac{\partial^2f}{\partial y^2}-2C_xC_y\frac{\partial f}{\partial x}\frac{\partial f}{\partial y}\frac{\partial^2 f}{\partial x\partial y}+[1+C_y(\frac{\partial f}{\partial y})^2]C_x\frac{\partial^2f}{\partial x^2}}{[1+C_x(\frac{\partial f}{\partial y})^2+C_y(\frac{\partial f}{\partial y})^2]^{3/2}}$$
Such method has been widely applied in the ARPES data analysis for a direct visualization of ARPES dispersions. For quantitative analysis, the dispersion is extracted by fitting the peak positions in EDCs (see e.g. Fig.~2g,h),  or momentum distribution curves (MDCs).  

\begin{figure}[H]
	\centering
	\includegraphics[width=16.8cm]{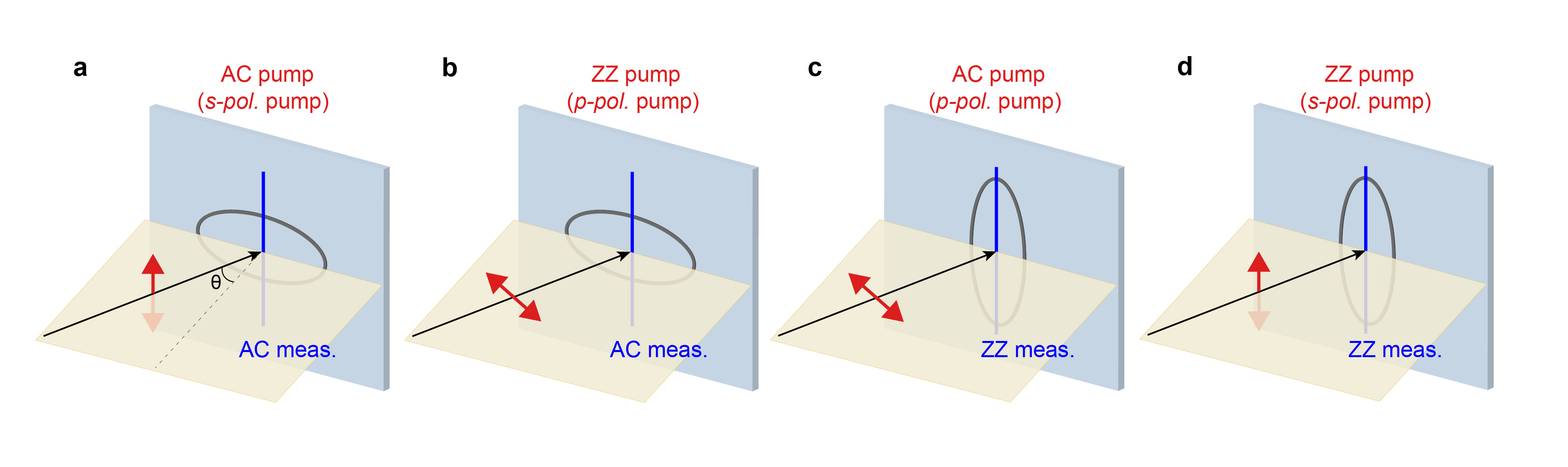}
	\caption{\textbf{Experimental geometries of TrARPES measurements.} \textbf{a-d}, Schematics for four different combinations of pump polarizations and measurement directions along AC or ZZ direction. The blue line indicates the  ARPES measurement direction and the red arrow represents the pump polarization. The black ellipses  represent the anisotropic electronic pocket of black  phosphorus.}
	\end{figure}

\subsection{More pump fluence dependent TrARPES data and analysis}
~\\
More pump fluence dependent TrARPES dispersion images are shown in Extended Data Fig.~2a-f. Clear gap opening is observed at high pump fluence (as pointed by the blue arrows). The size of the hybridization gap is extracted from the zoom-in EDCs at the resonance momentum positions shown in Extended Data Fig.~2h-m, and the extracted values are plotted in Extended Data Fig.~2n as a function of pump fluence.  The extracted light-induced gap shows a scaling of $\Delta \sim F^{0.5}$ with the pump fluence $F$, supporting the Floquet engineering interpretation of the light-induced band renormalization.

\begin{figure}[H]
	\centering
	\includegraphics[width=16.8cm]{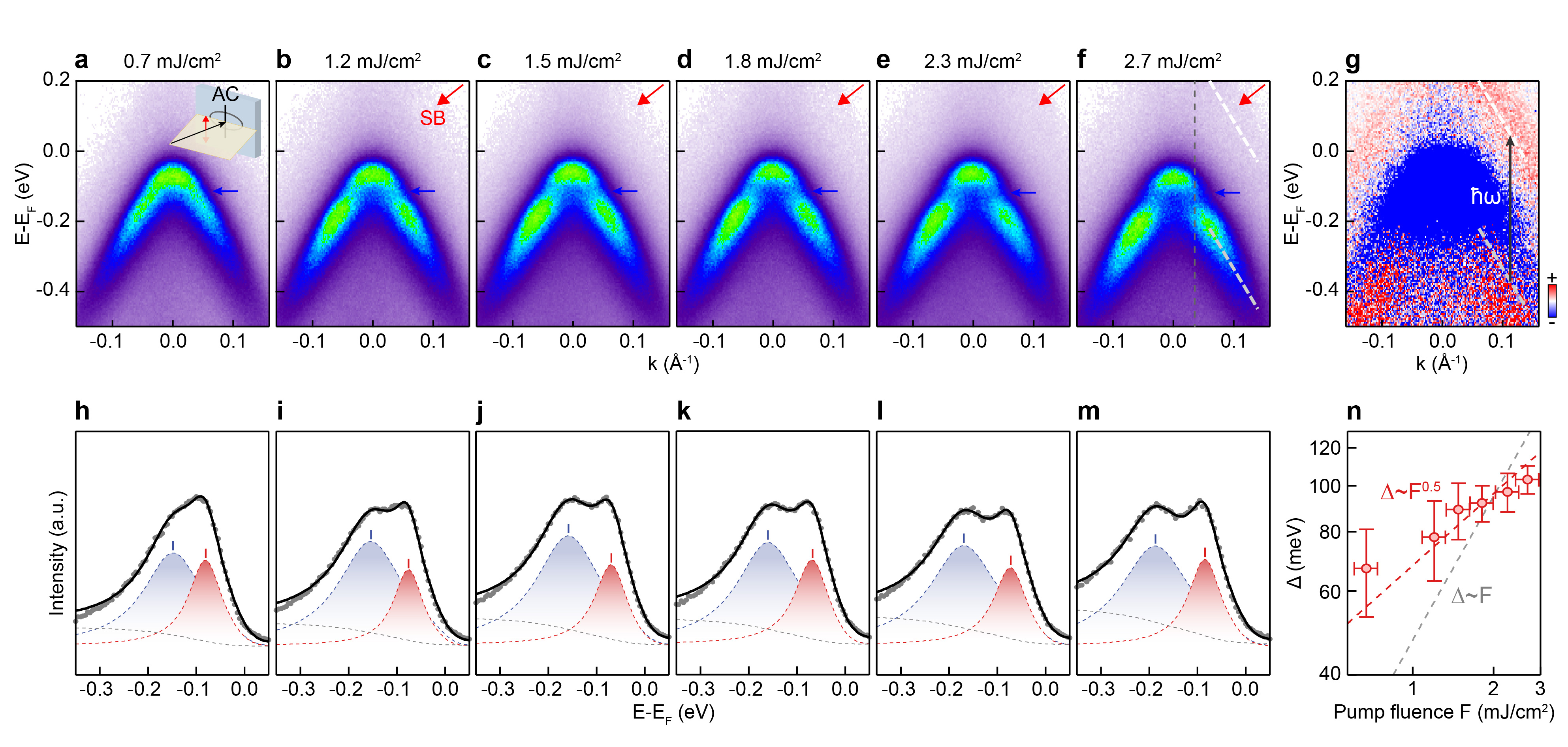}
	\caption{\textbf{Pump fluence dependent hybridization gap and observation of Floquet sidebands.} {\bf a-f}, TrARPES dispersions along the AC direction with different pump fluences ($s$-$pol.$ pump, $\hbar \omega$ = 420 meV, $\Delta$t = 0). The red and blue arrows point to the Floquet sideband of VB and hybridization gap. {\bf g}, The differential TrARPES dispersion with pump fluence of 2.7 mJ/cm$^2$ after subtracting the dispersion before pump.  {\bf h-m}, EDCs for data shown in {\bf a-f} at the momentum point of the hybridization gap (the dashed line in {\bf f}) and fitting curves.  \textbf{n}, The extracted hybridization gap as a function of pump fluence. The error bar of hybridization gap is defined as the upper limit when the energy position is clearly offset from the peak in \textbf{h-m},  and the error bar of pump fluence is defined by the fluctuation of pump power (10$\%$). }
	\end{figure}

\subsection{Exclusion of alternative explanation based on charge depletion}
~\\
To check if charge depletion can explain our experimental results, TrARPES measurements are performed on {\it in situ} electron-doped sample, so that the CB and VB near the band edges are both below the Fermi energy.  The fact that these states are both occupied means that optical transition upon near-resonance pumping is forbidden.  

A comparison of near-resonance pumping on undoped sample (Extended Data Fig.~3a-f) versus doped sample (Extended Data Fig.~3g-l) shows that although optical transition is blocked in electron-doped black phosphorus (see schematic illustration in Extended Data Fig.~3g), the light-induced gap is still clearly observed (Extended Data Fig.~3i-l), thereby indicating that the light-induced gap is not related to charge depletion upon optical absorption. Therefore, the observation of similar light-induced renormalization upon near-resonance pumping, regardless of when direct optical transition is allowed or forbidden indicates that it is not caused by charge depletion, but rather by the Floquet band engineering.

\begin{figure}[H]
	\centering
	\includegraphics[width=16.8cm]{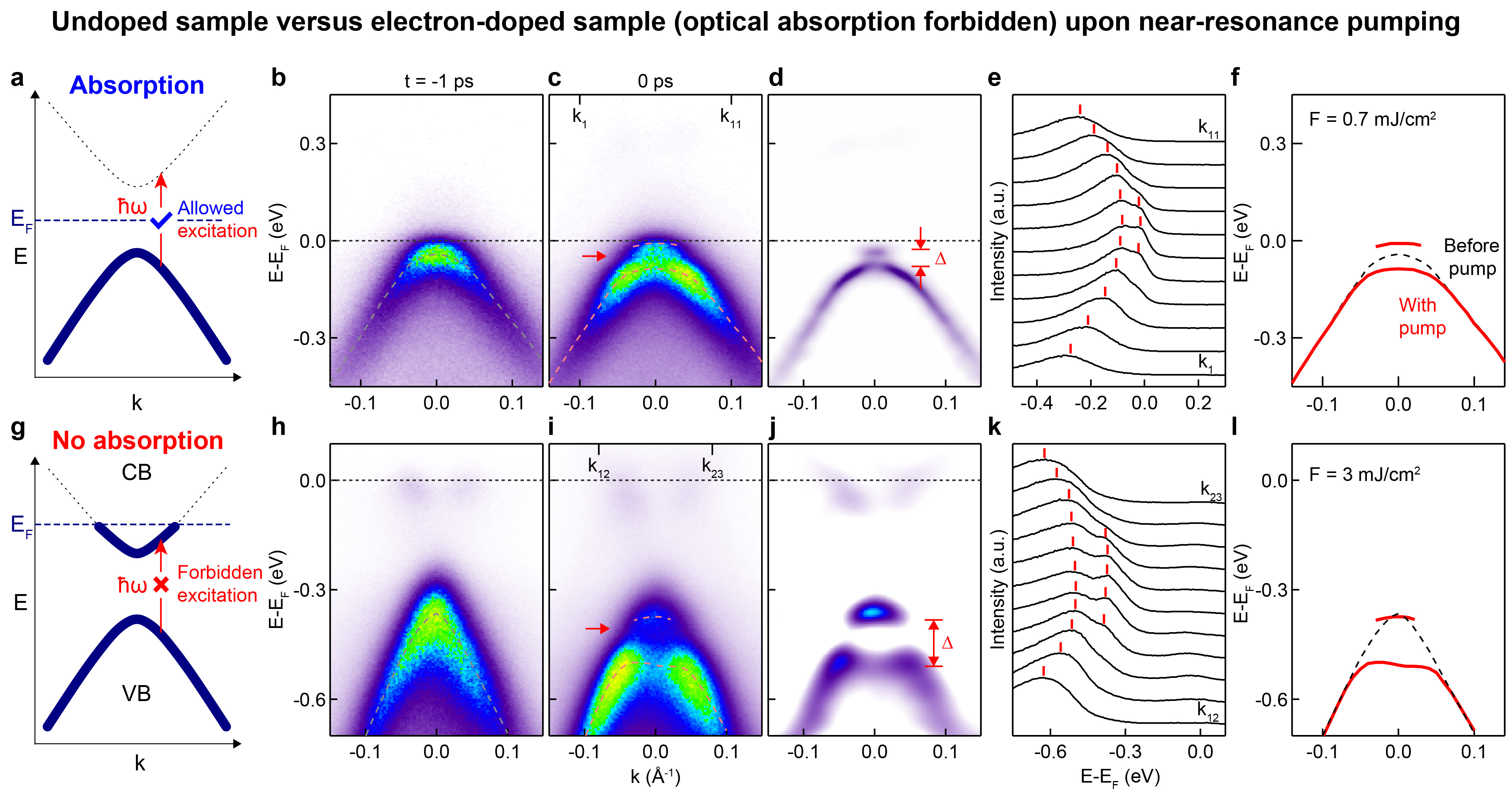}
	\caption{\textbf{Undoped sample versus electron-doped sample (forbidden optical absorption) upon near-resonance pumping.} {\bf a}, Schematic illustration of the electronic structure of undoped black phosphorus with unoccupied CB. Direct optical transition from VB to CB is allowed upon near-resonance pumping. {\bf b,c}, TrARPES dispersion images measured in undoped sample along AC direction at $\Delta$t = -1 ps (\textbf{b}) and $\Delta$t = 0 (\textbf{c}). The pump beam is polarized along AC direction with photon energy of 380 meV, and the pump fluence is 0.7 mJ/cm$\rm^2$. {\bf d,e}, Second derivative image and EDCs of TrARPES data shown in {\bf c}. {\bf f}, Extracted dispersions before pump (black dashed curve) and with pump (red curves). {\bf g}, Schematic illustration of the electronic structure of electron-doped black phosphorus with occupied CB edge. Direct optical transition from VB to CB is forbidden upon near-resonance pumping. {\bf h-l}, Similar results as {\bf b-f} but in electron-doped sample using pump pulses with photon energy of 400 meV and pump fluence of 3 mJ/cm$\rm^2$. The Fermi energy is obtained by the Fermi-Dirac fitting to the spectrum of electron-doped black phosphorus before pump ($\Delta$t = -1 ps).}
\end{figure}

\subsection{Comparison of experimental results for AC pump with different scattering geometries}
~\\
The pump polarization with respect to the scattering plane ($p$-$pol.$ or $s$-$pol.$) can lead to selective excitation of Floquet-Volkov states versus pure Floquet states\cite{mahmood2016}. For black phosphorus, the sideband shows a stronger intensity for $p$-$pol.$ pump due to Floquet-Volkov states and a weaker intensity for $s$-$pol.$ pump due to pure Floquet states, which is consistent with reported results on Bi$_2$Se$_3$ (ref.~\cite{mahmood2016}). More importantly, the band renormalization (pointed by red arrows in Extended Data Fig.~4) and sideband are observed for both $p$-$pol.$ and $s$-$pol.$ pump scattering geometries, similar to the case of Bi$_2$Se$_3$, suggesting that the light-induced band renormalization is a more intrinsic effect of Floquet engineering than the sideband intensity. Such band renormalization is important evidence to distinguish the Floquet band engineering from simply the emergence of Volkov states, because Volkov states do not renormalize the electronic structure of the host quantum materials, while the Floquet engineering does. 

\begin{figure}[H]
	\centering
	\includegraphics[width=16.8cm]{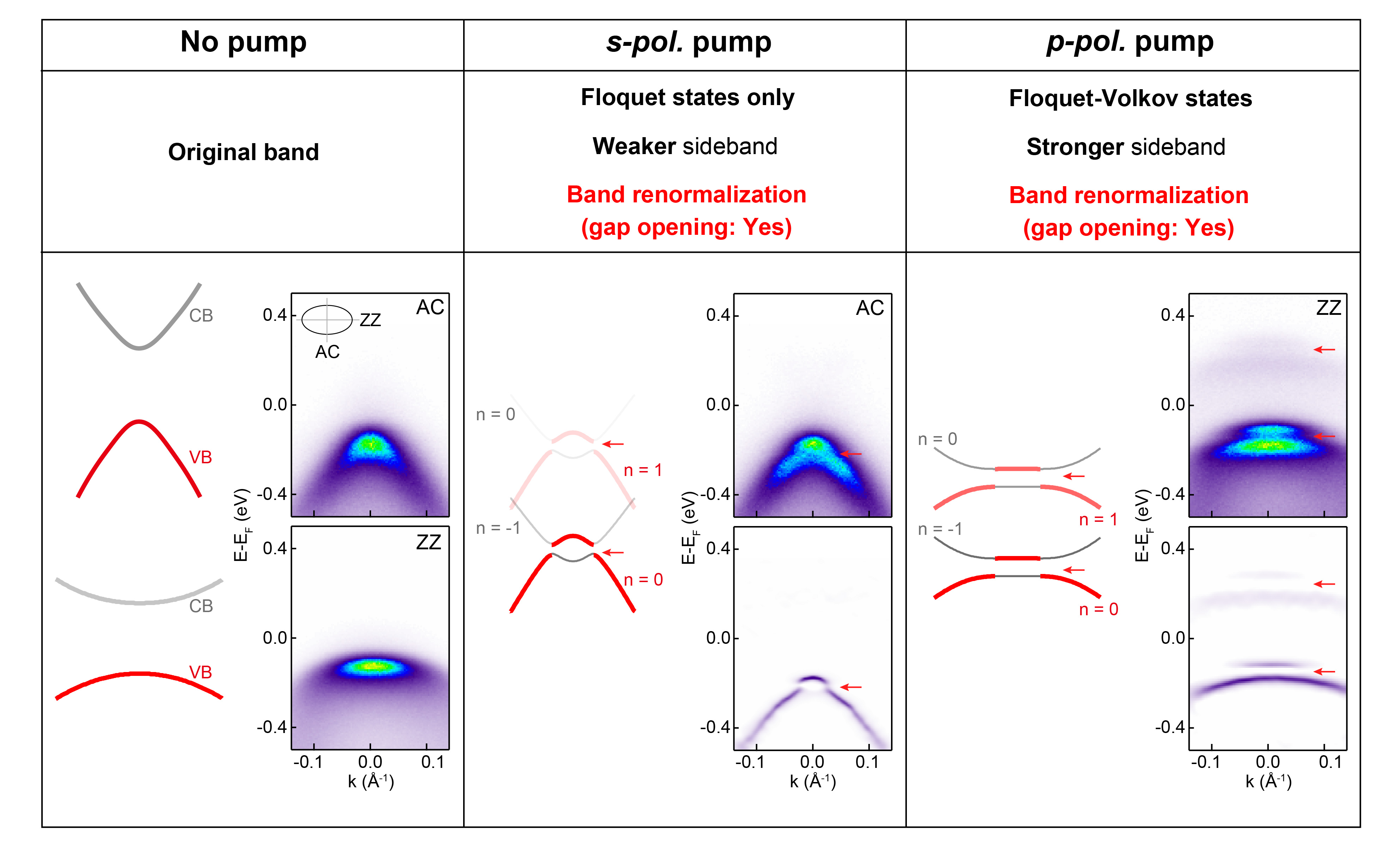}
	\caption{\textbf{Summary of Floquet band engineering for black phosphorus for $s$-$pol.$ pump and $p$-$pol.$ pump when the pump polarization is parallel to AC direction.} The pump photon energy is 440 and 380 meV for $s$-$pol.$ and $p$-$pol.$ pump respectively. The pump fluence is 0.7 mJ/cm$\rm^2$.}
\end{figure}

\subsection{Pump polarization dependent Floquet band engineering using the tight-binding calculation and $\bm k\cdot \bm p$ model}
~\\
Through the analysis of the tight-binding calculation and the Floquet $\bm k\cdot \bm p$ Hamiltonian around the $\Gamma$ point (see more details in the Supplementary Information), we find that the pseudospin selection rules in the equilibrium electronic state in black phosphorus, which are constrained by the lattice symmetry, strongly influence its hybridization gaps in the Floquet electronic structures. A large hybridization gap is observed along both AC and ZZ directions only for AC pump as shown in Extended Data Fig.~5. 

\begin{figure}[H]
	\centering
	\includegraphics[width=16.8cm]{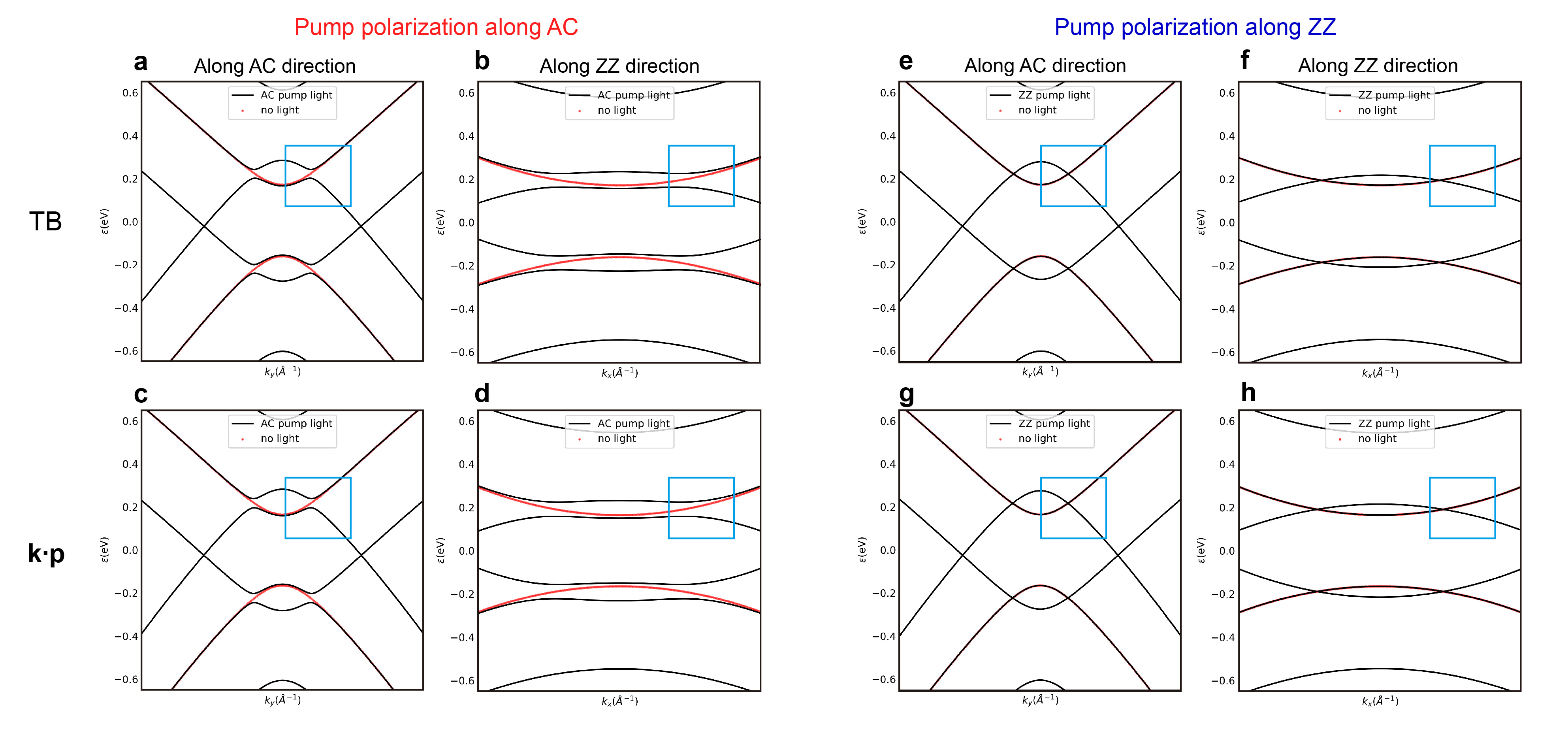}
	\caption{\textbf{Theoretical calculation of Floquet band structure.} \textbf{a-d}, Floquet band structure with AC pump for dispersions along AC and ZZ direction using tight-binding model ({\bf a,b}) and  \bm{$k \cdot p$} model ({\bf c,d}). \textbf{e-h}, Similar results as {\bf a-d} but with ZZ pump.}
\end{figure}

\subsection{Pump photon energy dependent Floquet band engineering and theoretical calculations}
~\\	
	The dispersions for different pump photon energies are calculated for dispersions along both AC (Extended Data Fig.~6) and ZZ (Extended Data Fig.~7) directions  from the $ab~initio$ tight-binding calculations within the Floquet theory, which are overall in good agreement with the TrARPES dispersions measured at $\Delta$t = 0. The hybridization gap is maximum upon near-resonance pumping.
	\begin{figure}[H]
		\centering
		\includegraphics[width=16.8cm]{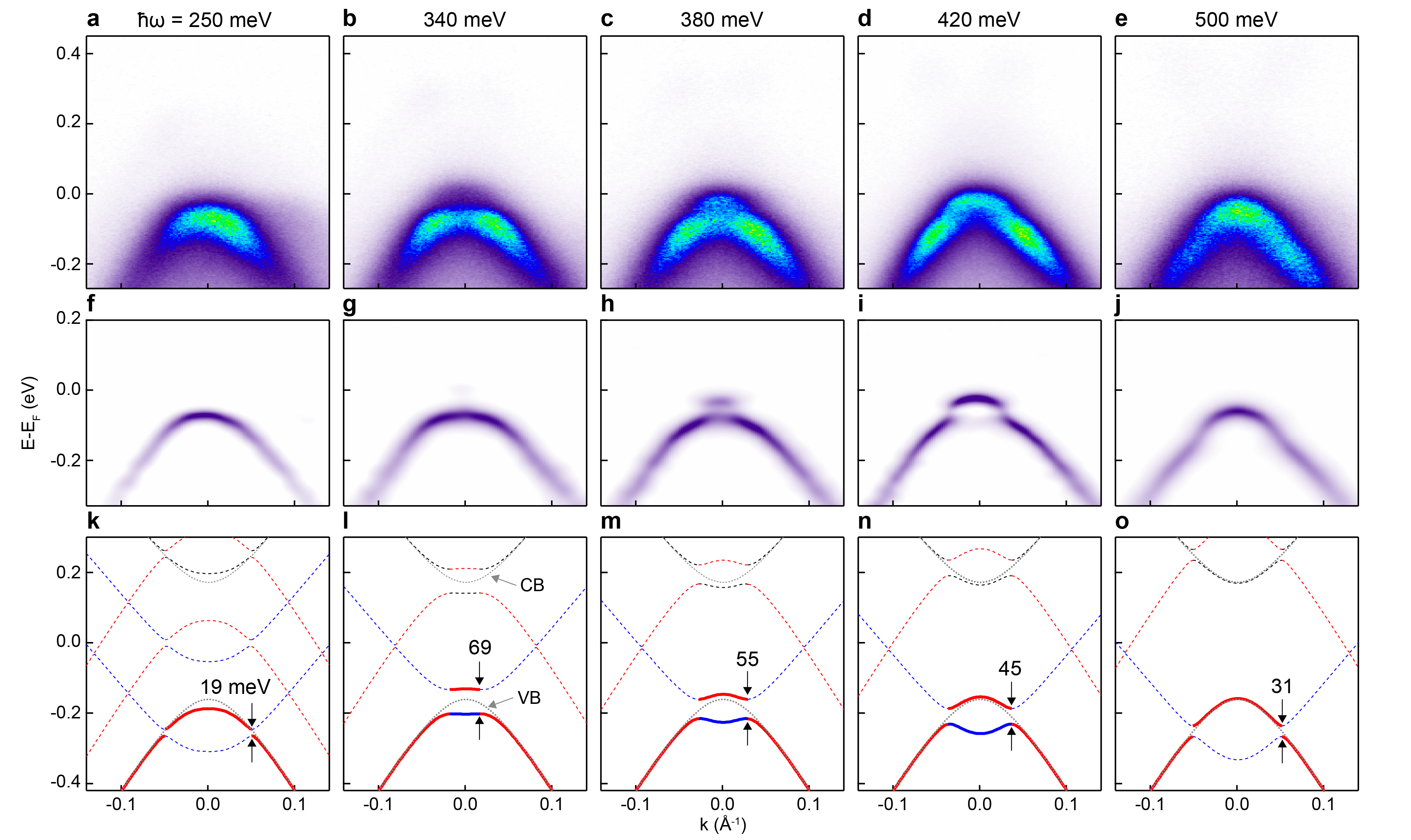}
		\caption{\textbf{Comparison of experimental and theoretical dispersions along AC direction for TrARPES with different pump photon energies.} \textbf{a-e}, TrARPES dispersions along AC direction (\textbf{a-e}) and corresponding second derivative images (\textbf{f-j}) with different pump photon energies at pump fluence of 0.7 mJ/cm$\rm^2$. \textbf{k-o}, Calculated Floquet band structures obtained from the $ab~initio$ tight-binding calculations with the Floquet theory to compare with \textbf{a-e}. The black dotted curves for the dispersions for the VB and CB in the equilibrium state for comparison.}
	\end{figure}

	\begin{figure}[H]
		\centering
		\includegraphics[width=16.8cm]{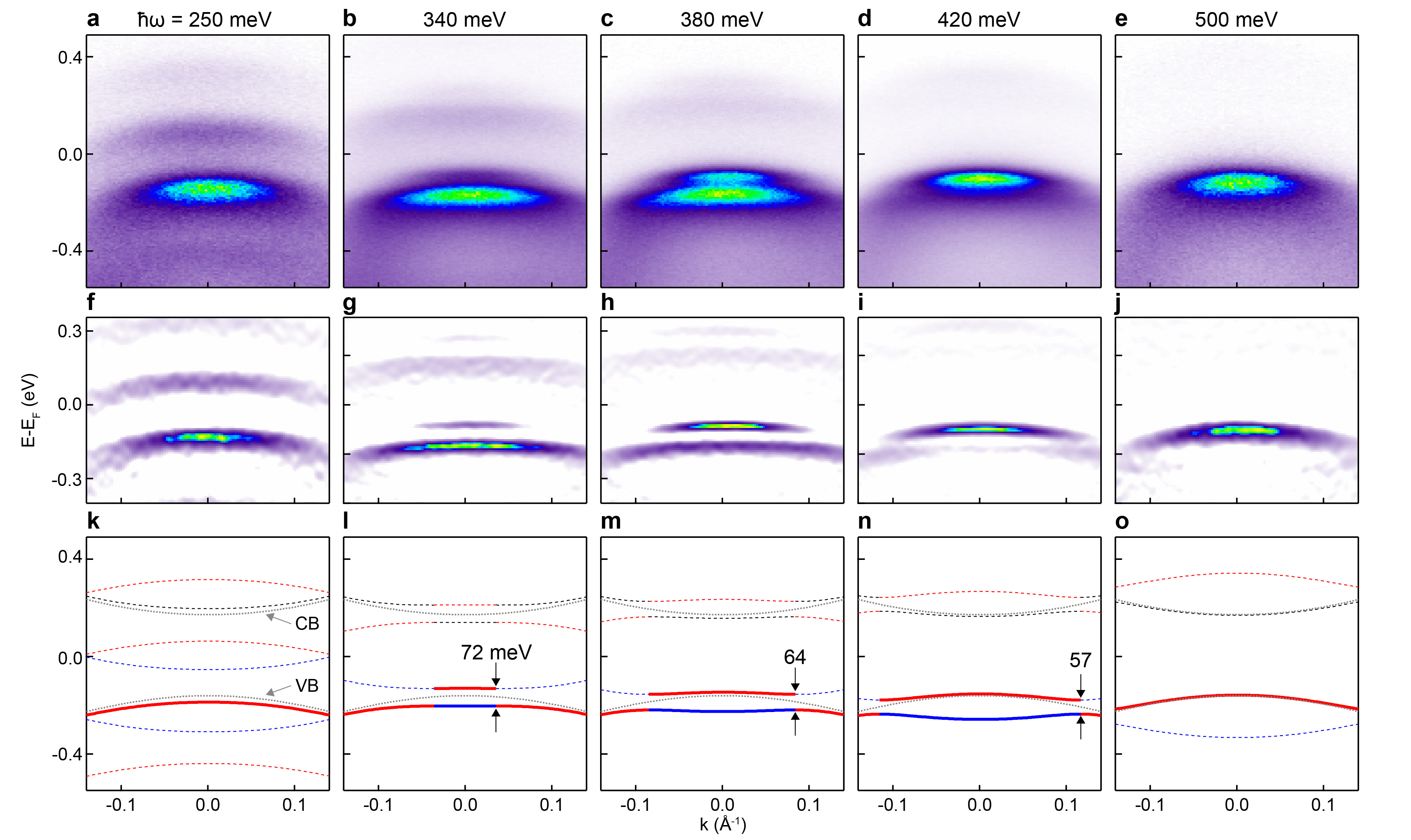}
		\caption{\textbf{Comparison of experimental and theoretical dispersions along ZZ direction for TrARPES  with different pump photon energies.} \textbf{a-j}, TrARPES dispersions along ZZ direction (\textbf{a-e}) and corresponding second derivative images (\textbf{f-j}) with different pump photon energies at pump fluence of 0.7 mJ/cm$\rm^2$. \textbf{k-o}, Calculated Floquet band structures obtained from the $ab~initio$ tight-binding calculations with the Floquet theory to compare with \textbf{a-e}. The black dotted curves for the dispersions for the VB and CB in the equilibrium state for comparison.}
	\end{figure}

\subsection{First-principles calculation}
~\\
	We performed the density functional theory (DFT) to calculate the electronic structure of black phosphorus without the laser pumping by using the Vienna $Ab~initio$ Simulation Package (VASP)\cite{PhysRevB.54.11169}.

	The projector augmented wave potentials\cite{PhysRevB.59.1758} were used and the cutoff of plane-wave energy was set as 400 eV, and the $\Gamma$ centered \emph{k}-mesh was sampled as $12\times12\times12$ in the Brillouin zone (BZ) of the primitive cell (see Extended Data Fig.~8a-c). The convergence condition of electronic self-consistent loop was $10^{-6}$ eV and the force criterion was set as 0.01 eV/\textrm{\AA}. The van der Waals corrections\cite{PhysRevLett.92.246401,PhysRevB.83.195131} were included both for lattice relaxation and electronic self-consistent calculations. We used the Perdew-Burke-Ernzerhof (PBE) type exchange-correlation functional\cite{PhysRevLett.77.3865} to obtain the optimized lattice structures. For the calculation of electronic state, we used the Heyd-Scuseria-Ernzerhof (HSE) hybrid functional\cite{Krukau2006InfluenceOT} to obtain the direct band gap as 330 meV at the Z point, which is consistent with our ARPES experimental observations. By using the wannier90 \cite{MOSTOFI2008685,MOSTOFI20142309,RevModPhys.84.1419}, we constructed the tight-binding Hamiltonian $\hat{H}^{TB}(\mathbf{r, k})$ from the $ab~initio$ calculations (see Extended Data Fig.~8d-f).

	\begin{figure}[H]
		\centering
		\includegraphics[width=16.8cm]{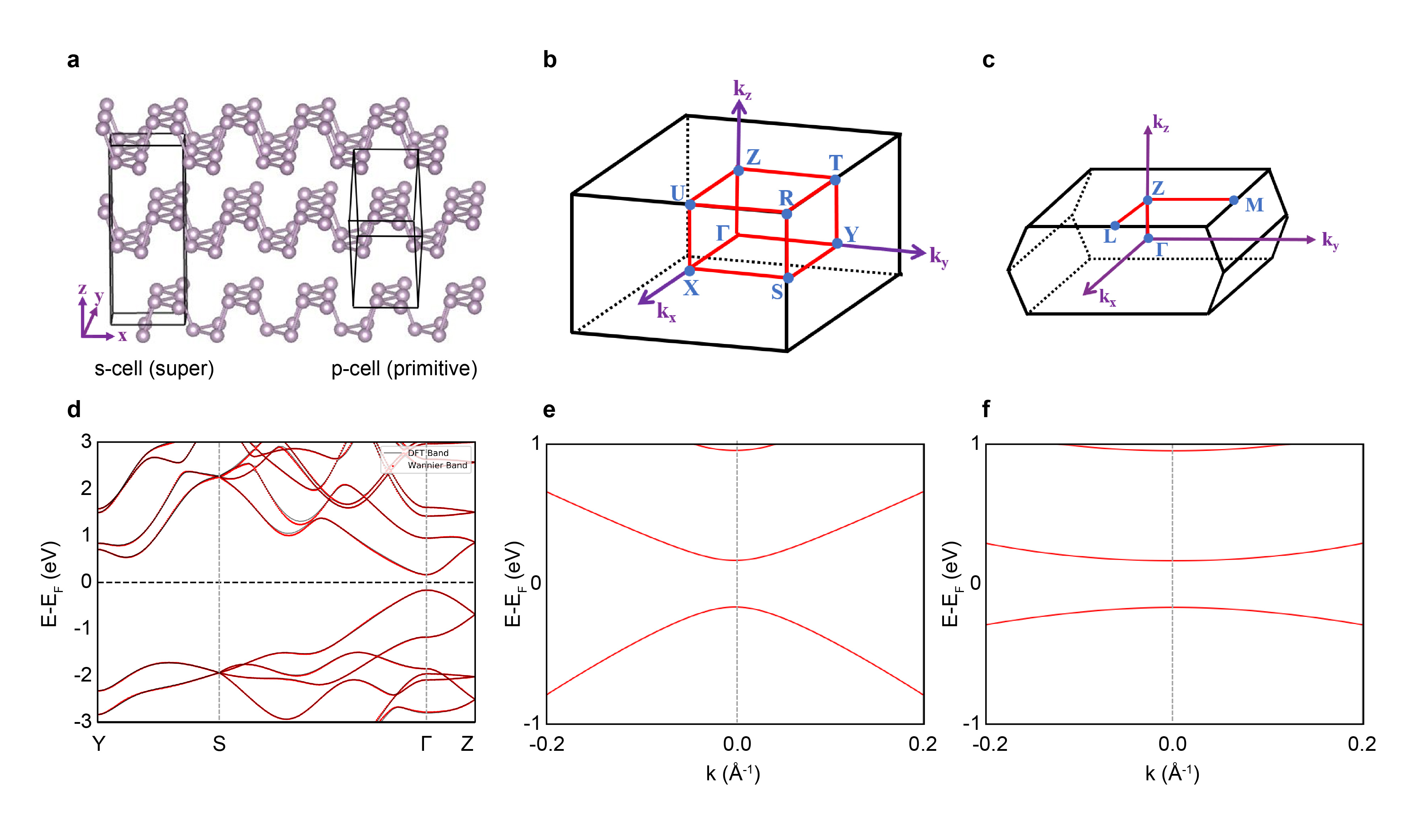}
		\caption{\textbf{Atomic and electronic structure of black phosphorus.} \textbf{a}, Atomic structure of black phosphorus. The primitive cell of 4 atoms is on the right and the supercell of 8 atoms is on the left.~\textbf{b}, The Brillouin zone (BZ) of the supercell. \textbf{c}, The BZ of the primitive cell. \textbf{d}, Comparison of the calculated band structures from DFT calculations and tight-binding calculations along high symmetry lines in the reduced BZ of supercell. \textbf{e,f}, The calculated band structure around the $\Gamma$ point along AC direction (\textbf{e}) and along ZZ direction (\textbf{f}). The Fermi level is set as zero.}
	\end{figure}

\subsection{Floquet Hamiltonian}
~\\
	We apply the Floquet theory to calculate the band structures of black phosphorus under the laser pumping. In our TrARPES experiments, the linear polarized probe pulse is used with the duration of 100 fs, which roughly corresponded to 10 optical cycles of pump pulse. In the calculations, we assume that the non-equilibrium Floquet band could be formed in a few optical cycles, which is confirmed by previous calculations via time-dependent DFT\cite{RubioNL2016}. Therefore, the Floquet theory is a good approximation to study the laser pumping effect in this case. Herein, we use the Peierls substitution to include the influence of light field, we obtain the time-dependent tight-binding Hamiltonian as
	\begin{equation}
		\hat{H}^{TB}(\mathbf{k}) \rightarrow \hat{H}^{TB}(\mathbf{k}+\frac{e}{\hbar}\mathbf{A}(t))
	\end{equation}
	where $\mathbf{A}(t)=(A_0\sin\Omega t, 0, 0)$ {or} $(0, A_0\sin\Omega t, 0)$ is the vector potential of pump light and $\Omega$ is its frequency. So the evolution of the wave-function $\Psi_{\alpha} (t)$ in the black phosphorus satisfies the time-dependent Schr{\"o}dinger equation
	\begin{equation}
		\label{H}
		\hat{H}^{TB}(\mathbf{k}, t)\left|\Psi_\alpha (t)\right\rangle = i \frac{\partial}{\partial t} \left|\Psi_\alpha (t)\right\rangle
	\end{equation}
	From the Floquet theorem, we could expand $\Psi_{\alpha} (t)$ in the Floquet basis  $\left|\Phi_{i}(t)\right\rangle$, it satisfies
	\begin{equation}
		\begin{aligned}
			\label{phi}
			&\left|\Phi_{\alpha}(t+T)\right\rangle=\left|\Phi_{\alpha}(t)\right\rangle \\
			&\left|\Psi_{\alpha}(t)\right\rangle=e^{-i \epsilon_{\alpha} t}\left|\Phi_{\alpha}(t)\right\rangle \\
		\end{aligned}
	\end{equation}
	where $\epsilon_{\alpha}$ are known as Floquet quasienergies and $T=\frac{2\pi}{\Omega}$. Now expand $\left|\Phi_{\alpha}(t)\right\rangle$ by a complete set of $\{\left|u^m_{\alpha}\right\rangle\}$
	\begin{equation}
		\label{psi}
		\left|\Phi_{\alpha}(t)\right\rangle=\sum_{m} e^{-i m \Omega t}\left|u_{\alpha}^{m}\right\rangle
	\end{equation}
	Bring Eq.(\ref{phi}) and Eq.(\ref{psi}) into Eq.(\ref{H}), we get
	\begin{equation}
		\sum_{m} \hat{H}^{TB}(\mathbf{k}, t) e^{-i m \Omega t}\left|u_{\alpha}^{m}\right\rangle=\sum_{m}\left[\epsilon_{\alpha}+m \Omega\right] e^{-i m \Omega t}\left|u_{\alpha}^{m}\right\rangle
	\end{equation}
	Finally we integrate against $e^{in\Omega t}$ on both sides of the above equation and obtain
	\begin{equation}
		\sum_{m}\frac{1}{T}\int_{0}^{T}dt\hat{H}^{TB}(\mathbf{k}, t) e^{i (n-m) \Omega t}\left|u_{\alpha}^{m}\right\rangle=\left[\epsilon_{\alpha}+n \Omega\right]\left|u_{\alpha}^{n}\right\rangle
	\end{equation}
	Let $\hat{H}^{FTB}_{n,m}(\mathbf{k})=\frac{1}{T}\int_{0}^{T}dt\hat{H}^{TB}(\mathbf{k}, t) e^{i (n-m) \Omega t}-m\Omega\delta_{mn}$, finally we get the Floquet tight-binding Hamiltonian satisfies the equation
	\begin{equation}
		\sum_{m}\hat{H}^{FTB}_{n,m}(\mathbf{k})\left|u_{\alpha}^{m}\right\rangle=\epsilon_{\alpha}\left|u_{\alpha}^{n}\right\rangle
	\end{equation}
	The index $\alpha$ labels eigenstates and $m$ and $n$ are the Fourier mode indices. Here, the time-dependent Schr{\"o}dinger equation is mapped to a time-independent eigenvalue problem in an extended Hilbert space. The Floquet Hamiltonian $\hat{H}^{FTB}(\mathbf{k})$ can be written as a block matrix
	\begin{equation}
		\hat{H}^{FTB}(\mathbf{k})=\left[\begin{array}{ccccc}
			\ddots & \vdots & \vdots & \vdots & \iddots \\
			\cdots & H^{FTB}_{-1,-1}(\mathbf{k})+\Omega I & H^{FTB}_{-1,0}(\mathbf{k}) & H^{FTB}_{-1,1}(\mathbf{k}) & \cdots \\
			\cdots & H^{FTB}_{0,-1}(\mathbf{k}) & H^{FTB}_{0,0}(\mathbf{k}) & H^{FTB}_{0,1}(\mathbf{k}) & \cdots \\
			\cdots & H^{FTB}_{1,-1}(\mathbf{k})& H^{FTB}_{1,0}(\mathbf{k}) & H^{FTB}_{1,1}(\mathbf{k})-\Omega I & \cdots \\
			\iddots & \vdots & \vdots & \vdots & \ddots
		\end{array}\right]
	\end{equation}
	We can diagonalize $\hat{H}^{FTB}(\mathbf{k})$ to obtain the Floquet band structure of the black phosphorus. The electric field of 6.8$\times 10^7$ V/m is used to calculate the TrARPES spectra.

\end{methods}

\begin{addendum}

 \item[Acknowledgements] We thank N.L. Wang, R.B. Liu, D. Sun and Y.H. Wang for useful discussions. This work is supported by the National Key R\&D Program of China (grant no.~2021YFA1400100), the National Natural Science Foundation of China (grant nos.~11725418, 12234011 and 11427903) and National Key R\&D Program of China (grant nos.~2020YFA0308800 and 2016YFA0301004). P.Y. and W.D. acknowledge the support of the Basic Science Center Project of NSFC (Grant No. 51788104). S.M. acknowledges supports from Ministry of Science and Technology (2021YFA1400201), National Natural Science Foundation of China (12025407) and Chinese Academy of Sciences (YSBR047).

 \item[Author contributions] Shuyun Z. conceived the research project. Shaohua Z., C.B., Q.G., Haoyuan Z. and T.L. performed the TrARPES measurements and analyzed the data. Haoyuan Z. grew the black phosphorus single crystal. B.F., Hui Z., H.L., P.T., S.M. and W.D. performed the theoretical analysis and calculation, and results shown in the manuscript are by B.F., P.T. and W.D.. C.B., Shaohua Z. and Shuyun Z. wrote the manuscript, and all authors contributed to the discussions and commented on the manuscript.

 \item[Competing interests] The authors declare that they have no competing financial interests.

 \item[Supplementary information] is available for this paper.
 	
\subsection{Data availability.}
	The data that support the findings of this study are available from the corresponding author upon request.

\end{addendum}

\end{document}